\documentclass{article}

\usepackage[english]{babel}
\usepackage{amssymb,physics}
\usepackage[letterpaper,top=2cm,bottom=2cm,left=3cm,right=3cm,marginparwidth=1.75cm]{geometry}
\usepackage{authblk}
\usepackage{amsmath}
\usepackage{graphicx}
\usepackage[colorlinks=true, allcolors=blue]{hyperref}
\title{Emergence of fractal cosmic space from fractional quantum gravity}
\author{P. F. da Silva J{\'u}nior, E. W. de Oliveira Costa, and S. Jalalzadeh\thanks{shahram.jalalzadeh@ufpe.br}}
\affil{Departamento de F\'{i}sica, Universidade Federal de Pernambuco, Recife, PE, 50670-901, Brazil}

\begin{document}
\maketitle

\begin{abstract}
Based on Padmanabhan's theory, the spatial expansion of the Universe can be explained by the emergence of space as cosmic time progresses. To further explore this idea, we have developed fractional-fractal Friedmann and Raychaudhuri equations for an isotropic and homogeneous universe. Our analysis has also delved into how Padmanabhan's concept fits into the framework of fractional quantum gravity. Our research shows that a fractal horizon model strongly supports the validity of the emerging Universe paradigm and its connection to horizon thermodynamics. This study indicates early how the emergent gravity perspective might manifest in quantum gravity. By utilizing the fractional-fractal Friedmann and Raychaudhuri equations, we have established that the mainstream cosmology model can be justified without a dark matter component. As a result, the standard $\Lambda$CDM model has been reduced to $\Lambda$-Cold Baryonic Matter, which has significant implications for our understanding of the Universe.
\end{abstract}

\section{Introduction} \label{Introduction}

Quantum gravity constitutes a vast research area that envelops frameworks that consistently integrate gravitational force and quantum physics. On the one hand, specific approaches quantize gravity as a fundamental force or incorporate it in a unified theory of elementary interactions. In this scenario, new physics will likely surface at short ultraviolet (UV) scales on the order of the Planck length. On the other hand, quantum gravity may constitute an effective field theory with infrared (IR) corrections visible at large scales. In this scenario, the emphasis does not necessarily lie in incorporating these models in a complete fundamental theory; rather, it lies in phenomenology and IR deviations from general relativity (GR) that may be visible in near-future cosmological observations \cite{Belgacem:2020pdz}. This alternate technique proves particularly effective in resolving the issues of quantum gravity, a subject of ongoing research in theoretical physics. Therefore, to gain a complete understanding of quantum gravity, it's imperative to take into account both the UV and IR regimes. In order to address the shortcomings of quantum mechanics and GR, researchers have proposed various approaches, including string theory, loop quantum gravity, and asymptotic safety. These theories aim to provide a consistent framework for understanding the universe's behavior at both the most minor and largest scales. 

The ultimate objective of quantum gravity research is to merge the fundamental forces of nature and establish a quantum theory of gravity that can explain all observed phenomena. Quantum field theory is commonly regarded as non-local due to non-polynomial momentum functions in loop corrections to the bare propagator. The modifications to the quantum effective action are typically non-local in nature, which can leave an imprint in the IR region, particularly in the case of gravity. This IR non-locality defines a specific regime of a theory that is essentially local. On the other hand, UV non-locality pertains to fundamentally non-local theories at the classical level. In contrast to IR non-locality, UV non-locality is not a result of quantum corrections but rather an intrinsic property of the theory itself. Therefore, it is important to distinguish between these two types of non-locality in order to understand the behavior of quantum field theories better. In GR, fundamental non-locality is utilized to overcome the classical singularities, improve its renormalizability, and maintain unitarity (the absence of ghosts and probability conservation). String theory \cite{Zwiebach:2004tj}, non-local quantum gravity with exponential or asymptotically polynomial form factors \cite{Modesto:2011kw,Biswas:2011ar,2017IJ0M}, fractional quantum gravity \cite{Jalalzadeh:2020bqu,Rasouli:2021lgy,Jalalzadeh:2021gtq,Rasouli:2022bug,Jalalzadeh:2022uhl}, and multi-fractional field theories with fractional operators \cite{Calcagni:2016azd,Calcagni:2021ipd,Calcagni:2021ljs} are examples of UV non-local theories, whereas IR non-locality is realized by models with inverse powers of the d'Alembertian \cite{Belgacem:2020pdz,Barvinsky:2003kg}.

Numerous applications of fractional calculus in the fields of
gravity and cosmology have been identified, and current research in this area is actively ongoing. Fractional calculus has proven to be a valuable tool in addressing a range of issues related to gravitational forces and cosmological models
\cite{Garcia-Aspeitia:2022uxz,Shchigolev:2010vh,Shchigolev:2012rp,Shchigolev:2013jq,Calcagni:2013yqa,Shchigolev:2015rei,Calcagni:2016azd,Shchigolev:2021lbm,Jalalzadeh:2022uhl,Calcagni:2020ads,Calcagni:2021ipd,Calcagni:2021aap,Leon:2023iaq,Gonzalez:2023who,Socorro:2023ztq}. In addition, the stochastic gravitational-wave background in quantum gravity \cite{Calcagni:2020tvw}, gravitational-wave luminosity distance \cite{Calcagni:2019ngc}, inflation and cosmic microwave background (CMB) spectrum \cite{Rasouli:2022bug,Calcagni:2017via,Calcagni:2016ofu}, fractional action cosmology \cite{El-Nabulsi:2012wpc,El-Nabulsi:2015szp,Jamil:2011uj}, 
 fractional geodesic equation, discrete gravity \cite{El-Nabulsi:2013mma}, non-minimal coupling and chaotic inflation \cite{El-Nabulsi:2013mwa}, phantom cosmology with conformal coupling \cite{Rami:2015kha}, Ornstein--Uhlenbeck like fractional differential equation in cosmology \cite{El-Nabulsi:2016dsj}, fractional action cosmology with a variable order parameter \cite{El-Nabulsi:2017vmp}, wormholes in fractional action cosmology \cite{El-Nabulsi:2017jss} are other examples in the application of fractional calculus in GR and cosmology. New metrics were considered \cite{El-Nabulsi:2017rdu}, as well as some dark energy models in emergent, logamediate, and intermediate scenarios of the Universe \cite{Debnath2012,Debnath2013}. For instance, \cite{Shchigolev:2015rei,Shchigolev:2021lbm} found $\alpha=0.926$ (where $\alpha$ is the order of the Riemann--Liouville fractional integral).  In \cite{Shchigolev:2010vh,Shchigolev:2012rp,Shchigolev:2013jq} were obtained several exact solutions for cosmological models, which, since space-time is fractal, deviates sharply from the standard model \cite{Calcagni:2010bj,Calcagni:2009kc}. The Ref. \cite{Jalalzadeh:2022uhl} explore the interval $1\leq\alpha <2$ using Riesz's fractional derivative to obtain the non-boundary and tunneling wave functions for a closed de Sitter geometry. In Ref. \cite{Rasouli:2022bug}, the pre-inflation epoch is studied in the context of fractional quantum cosmology. The thermodynamics of fractional BHs has been studied in Ref. \cite{Jalalzadeh:2021gtq}.  
 Another example is the fractional calculus modification of the Friedmann and Raychaudhuri equations to study the dynamics of the Universe without the presence of cold dark matter (CDM) and dark energy \cite{Barrientos:2020kfp}. A further approach involves utilizing fractional calculus to determine the value of the cosmological constant, which must be restructured owing to the well-known ultraviolet divergence in traditional quantum field theory \cite{Landim:2021www,Calcagni_2021,Landim:2021ial}. 
 In \cite{Giusti:2020rul,Torres:2020xkw} explore modified Newtonian dynamics (MOND) and quantum cosmology in this fractional approach \cite{Barrientos:2020kfp}. Finally, notice that there are several definitions of fractional derivatives and fractional integrals, such as those of Riemann--Liouville, Caputo, Riesz, Hadamard, Marchand, and Griinwald--Letnikov, among other more recent ones (see \cite{book1:2006}, and \cite{book2:1999}  and references therein). Even though these operators are already well studied, some of the usual features related to function differentiation fail, such as Leibniz's rule, the chain rule, and the semi-group property \cite{book2:1999,book1:2006}.

 According to a recent study conducted by the authors of Ref. \cite{Jalalzadeh:2021gtq}, the surface area of a Schwarzschild black hole (BH) can be considered a random fractal due to a modification made to the horizon area using the fractional Wheeler--Dewitt equation. This new development changes how the entropy of BHs is calculated, as it no longer follows the traditional Bekenstein--Hawking area law and may increase due to the quantum gravitational effects. The formula used to calculate the entropy of the BH is now expressed as
 \begin{equation*}
     S_\text{fractal-BH}=S_\text{BH}^\frac{d}{2}
 \end{equation*} 
 where $S_\text{BH}=4\pi GM^2$ represents the Bekenstein--Hawking entropy of a Schwarzschild BH and $d$ is the fractal dimension of the BH surface. 
The aforementioned expression for fractionally deformed entropy may bear some resemblance to Tsallis \cite{2013EPJCT}, and Barrow \cite{Barrow:2020tzx} entropies, yet it is important to note that its correction, motivation, and physical principles are altogether separate and unique.

The main objective of this study is to evaluate the efficiency of Padmanabhan's emerging cosmic space model \cite{Padmanabhan:2012ik} using fractional entropy developed within the framework of fractional quantum cosmology (FQC). FQC has been introduced in different references, including \cite{Rasouli:2022bug,Jalalzadeh:2022uhl,Jalalzadeh:2021gtq,Rasouli:2021lgy,Moniz:2020emn}, and provides a unique framework for examining the structure of the Universe. By employing this strategy, we seek to acquire insight into the Universe's matter composition and improve our grasp of its fundamental features. The birth of Padmanabhan’s emergent cosmic space paradigm
lies in the connection between thermodynamics and gravity. In the context of BHs, the term ''surface gravity'' refers to the temperature where the temperature points are on the BH’s horizon.
However, the BH’s entropy is also related to this horizon’s region. With that, a solid relationship between thermodynamics
and gravity was not discovered until Jacobson’s investigations. 
 Jacobson's work began with the Clausius equation, $dQ= TdS$, from which, using the equivalence principle, Einstein's field equations were constructed, demonstrating that they constitute the equation of state for space-time \cite{Jacobson:1995ab}.

Padmanabhan postulated \cite{Padmanabhan:2012ik} that the variation of the cosmic volume, $dV$, in an infinitesimal interval of cosmic time, $dt$, is given by
\begin{equation*}
       \frac{dV}{dt}=L_\text{P}^2(N_\text{sur}-N_\text{bulk}),
\end{equation*}
where $N_\text{sur}$ and $N_\text{bulk}$ are the surface and bulk degrees of freedoms, respectively. While the surface degree of freedom is proportional to the Universe's horizon surface, the bulk degree of freedom is proportional to the Komar energy and inversely to the horizon's temperature. By combining Padmanabhan's equation with the continuity equation for the matter field, which is often assumed to be a perfect fluid, we can obtain Friedmann and Raychaudhuri's equations.

The model proposed by Padmanabhan has sparked a great deal of interest among scholars who have conducted further investigations into its potential generalizations. For instance, in the study by Cai in 2012 \cite{Cai:2012ip}, the original model was extended by modifying the volume increase and the number of degrees of freedom on the holographic surface from the entropy formulas of black holes in Gauss--Bonnet gravity and more general Lovelock gravity. Similarly, a paper by Tu and colleagues in 2013 \cite{Tu:2013gna} discovered a general relationship between the horizon entropy and the number of degrees of freedom on the surface, which can be applied to quantum gravity. The corresponding dynamic equations were obtained using the idea of the emergence of spaces in the $f(R)$ theory and deformed Hořava--Lifshitz theory. In their 2013 publication \cite{Hashemi:2013oia}, Hashemi and colleagues demonstrated that the apparent horizon is the only horizon to which all thermodynamic laws apply. They also provided a set of cosmology equations for information and thermodynamical parameters in their 2015 paper \cite{Hashemi:2015vsa}, including a generalized form for the Bekenstein--Hawking entropy for both the holographic principle and the asymptotic holographic principle. In 2013, Yuan \cite{Yuan:2013nsa} investigated the cases where logarithmic and power-law entropic corrections are present, respectively. Additionally, Ali \cite{Ali:2013qza} derived a modified Friedmann equation in Gauss--Bonnet gravity by considering a generic form of entropy as a function of the area in their 2013 study. Moradpour's 2016 paper used Tsallis entropy to bridge Verlinde’s and Padmanabhan’s proposals \cite{Moradpour:2016rcy}. Padmanabhan’s proposal was also exploited to study the Rastall theory by extending the Komar energy and the general entropy of the apparent horizon in Yuan's 2016 publication \cite{Yuan:2016pkz}. In 2016, Komatsu and colleagues applied a modified Rényi entropy to Padmanabhan’s holographic equipartition law by regarding the Bekenstein--Hawking entropy as a non-extensive Tsallis entropy and using a logarithmic formula of the original Rényi entropy \cite{Komatsu:2016vof}. Finally, in studies by Sheykhi and Chen in 2018 and 2022 \cite{Sheykhi:2018dpn,Chen:2022ymq}, respectively, Padmanabhan's emergence scenario of the modified Friedmann equations was employed utilizing Tsallis entropy. Barrow entropy was also employed in studies by Sheykhi and Luciano in 2021, and 2023 \cite{Sheykhi:2021fwh,Luciano:2023zrx}, respectively, to extend Padmanabhan’s proposals.

A common and effective method utilized in all these articles is applying a generalized definition of entropy, originating from BH entropy, to redefine the effective area of the apparent horizon, the number of surface degrees of freedom, and the increase in the effective volume of the Universe. This is followed by applying Padmanabhan's emergence of gravity and the first law of thermodynamics to arrive at the modified Friedmann equation. In this article, we take a comparable approach to the references mentioned earlier. Our objective is to broaden the fractional entropy of the black hole and encompass the apparent horizon of a homogeneous and isotropic universe. To achieve this goal, we must redefine an effective form for all the constituents that are associated with Padmanabhan's emergence equation and the first law of thermodynamics. Eventually, this procedure leads us to the fractional-fractal \footnote{It's worth noting that a fractal can be described as an object or process with a fractal dimension. When fractional calculus is applied to such objects, it can alter their fractal dimension \cite{tatom1995}. This article explores this relationship and how the fractal structure emerges through the process of applying fractional calculus. To emphasize both qualities, we use the term ''fractal-fractional.''} Friedmann and Raychaudhuri equations.

Our present comprehension of the Universe is founded on the postulate that General Relativity (GR) and the standard model of particle physics accurately portray its underlying physics. It is believed that the large-scale geometry of the Universe is flat, while its material composition is intricate, comprising of baryonic matter, CDM, and dark energy. This standard cosmological model has been substantiated by gauging temperature and polarization fluctuations in the CMB from both space and ground-based observatories. Remarkably, a model requiring just six independent variables provides an exhaustive fit to all statistical characteristics of recent CMB measurements and corroborates with the distribution of galaxies, Hubble constant, and supernova distance measurements. Nevertheless, this triumph comes with a price: dark energy constitutes the majority of the Universe's energy density, while CDM accounts for less than 5\% of its ordinary matter content but governs the galaxy mass. Evidence from rotation curves, gravitational lensing, and hot gas in clusters suggests that 95\% of the mass of galaxies and clusters is formed of CDM. The gravitational clustering of CDM serves as the cornerstone for the current paradigm of structure formation, as baryonic matter alone is insufficient to produce structures compatible with galaxy clustering.

Meanwhile, as indicated above, all astronomical arguments of dark matter presuppose that GR holds true on galaxy scales. Alternatives theories of gravity, including MOND \cite{Milgrom:1983ca,Sanders:2002pf,Famaey:2011kh,Bekenstein:2004ne}, eliminate the necessity for CDM by altering the nature of gravity. The notion of using anomalous fractional dimensions to actualize CDM galaxy data has recently been revived, implicitly in Newtonian dynamics with a fractional Laplacian \cite{Giusti:2020rul,Giusti:2020kcv} and explicitly in Newtonian fractional-dimension gravity \cite{Varieschi:2020ioh,Varieschi:2020dnd,Varieschi:2020hvp,Calcagni:2021mmj}.

One intriguing concept to consider is expanding the fractal-fractional gravity models explained in the above paragraph to the cosmological level. Could these adjustments eliminate the necessity for CDM at this level if fractional-fractal dynamics can account for the flattening of galaxy rotation curves? We embrace Padmanabhan's paradigm and deduce the fractional-fractal Friedmann and Raychaudhuri equations to delve into this inquiry. By utilizing these equations, we recognize novel density parameters for dark energy and cold matter (or dust) that are altered by the fractal dimension of the cosmic event horizon. Our study demonstrates that the density parameter for fractal cold matter can substitute for the need for a CDM component. This parameter is created solely by baryonic matter in a fractal distribution of the Universe.

The following is an outline of this article. The essential ingredient of Padmanabhan’s paradigm is the cosmic horizon.
Therefore, the following section briefly reviews the fractional-fractal quantum BH obtained in Ref. \cite{Jalalzadeh:2021gtq}. Thereafter, in section 3, we review some
basics of fractals. This section is included to help readers
quickly pursue our definitions in the next section. Utilizing
definitions of section 3, we demonstrate how the Schwarzchild
BH’s event horizon is fractal structures in section 4. We
show that the BH’s horizon is a fractal surface with dimension $2\leq d<3$. The effective Schwarzschild radius, the BH's mass, area, temperature, and entropy are obtained in this section.
 After briefly reviewing emergent cosmology in section 5, we apply the fractal-fractional quantities developed in section 4 to the cosmic space emergence scenario to get the modified Raychaudhuri and Friedmann equations in section 6. After reviewing the basics of the standard $\Lambda$CDM cosmology in section 7, we analyzed its fractal-fractional extension in section 8 and demonstrated that the $\Lambda$CDM cosmology reduces to $\Lambda$-Cold Baryonic Matter. In the final section, we draw our conclusions.
We will refer to natural units in this
paper as $\hbar=c=k_B=1$, for convenience.

\section{fractional quantum gravity of Schwarzschild black hole}\label{Horizon}

In this section, we will provide a brief summary of the findings presented in \cite{Jalalzadeh:2021gtq}. This information will be used in upcoming sections. 
 
To achieve the desired fractional mass spectrum of a Schwarzschild BH, let us first briefly explain the fractional Wheeler--DeWitt (WDW) in a $v$-dimensional minisuperspace with coordinates $q^\alpha$, $(\alpha=0,...,v)$. In this case, the minisuperspace WDW equation is
\begin{equation}\label{WDW}
    \left\{\frac{1}{2}\Box+U(q^\sigma)\right\}\Psi(q^\eta)=0.
\end{equation}
In this equation, $\Box=\frac{1}{\sqrt{-f}}
\partial_\alpha(\sqrt{-f}f^{\alpha\beta}\partial_\beta)$ is the d’Alembertian operator in minisuperspace, $U(q^\nu)$ is potential, and
$f_{\alpha\beta}$ stands for the corresponding minisuperspace metric
with signature $(-,+,+,...,+)$.

Regarding obtaining the fractional counterpart of the
WDW equation \eqref{WDW}, the usual d'Alembertian operator should be replaced by the fractional Riesz-d'Alembertian operator
\cite{Rie,Tarasov:2018zjg}
\begin{equation}\label{Riesz-d'Alembertian}
(-\Box)^{\frac{\alpha}{2}} \Psi(q^\alpha)
={\mathcal F}^{-1}\Big(|{\boldsymbol p}|^\alpha{\mathcal F}\Psi({\boldsymbol p})\Big),
\end{equation}
where $|{\boldsymbol{p}}|=\sqrt{-p_0^2+p_ip^i}$, $i=1,2,...,v$, and ${\mathcal F}$ denotes Fourier transformation. Hence, the fractional counterpart of the WDW equation (\ref{WDW}) will be \cite{Jalalzadeh:2020bqu,Moniz:2020emn,Rasouli:2021lgy,Jalalzadeh:2021gtq,Jalalzadeh:2022uhl}
\begin{equation}\label{WDWsh}
    \left\{\frac{M_\text{P}^{2-\alpha}}{2}(-\Box)^\frac{\alpha}{2}-U(q^\nu)\right\}\Psi(q^\nu)=0,
\end{equation}
and the Lévy’s fractional parameter, $\alpha$,  is set to $1 <\alpha \leq 2$ \cite{Laskin:2002zz}.

Bekenstein postulated that a BH's entropy is proportional to its horizon area in the early 1970s of the previous century \cite{Bekenstein:1973ur}. Subsequently, Hawking proposed in 1975 the evaporation of BHs \cite{Hawking:1975vcx}, which recent appraisal that tested this property through observation produced encouraging findings \cite{Isi:2020tac}. After that, theoretical physicists discovered a close link between thermodynamic temperature, quantum physics, and geometrical horizons. It was also suggested that the BH horizon area,
$A$ might be quantized, and the associated eigenvalues would be \cite{Bekenstein:1974jk}
provided by
\begin{equation}
    \label{H1}
    A_n=\gamma{\,} L_\text{P}^2{\,}n,~~~~n=1,2,3,...~,
\end{equation}
where $\gamma$ is a dimensionless constant of order one and $L_\text{P}=\sqrt{G}$ is the Planck length. Since its inception, the realm of literature has been imbued with a plethora of contributions, all of which have served to reinforce the veracity of the area spectrum (\ref{H1}). These contributions, which are both multifaceted and diverse, encompass a broad range of considerations, including but not limited to information-theoretic considerations (as set forth in \cite{Danielsson:1993um,Bekenstein:1995ju}), string theory arguments \cite{Mazur:1987jf}, and the periodicity of time \cite{Mazur:1986gb,Mazur:1987sg,Bina:2010ir}. Furthermore, these contributions extend to a Hamiltonian quantization of a dust collapse \cite{Peleg:1995gg,Nambu:1987dh}, among other things.

It is well-known that the above spectrum, among the various methods, can be obtained by the canonical quantization of the Schwarzschild BH \cite{Louko:1996md}. To review this process, let us start with the Hamiltonian of the Schwarzschild BH given by \cite{Louko:1996md}
\begin{equation}
    \label{H2}
    H=\frac{M^2_\text{P}}{2}\left(\frac{p^2}{M_\text{P}^4{\,}x}+x\right)=M,
\end{equation}
where $M$ is the BH's mass,  $M_\text{P}=1/L_\text{P}$ denotes the Planck mass, and $p$ is conjugate momenta of $x$ constrained to condition $x\geq0$.
The Wheeler--DeWitt equation corresponds to the Schwarzschild BH \cite{Jalalzadeh:2021gtq}, is given by
\begin{equation}
    \label{H3}
    -\frac{1}{2M_\text{P}}\frac{d^2\psi(x)}{dx^2}+\frac{M_\text{P}^3}{2}\left(x-\frac{M}{M_\text{P}^2}\right)^2\psi(x)=\frac{M^2}{2M_\text{P}}\psi(x),
\end{equation}
which provides us with the following mass spectrum in the semi-classical limit
\begin{equation}
    \label{H4}
    M_n=M_\text{P}\sqrt{2n}, ~~~~n=\{\text{large positive integer}\}.
\end{equation}
Assume that a huge BH emits Hawking radiation when it spontaneously transitions from state $n+1$ to the nearest lower state level, i.e., $n$. Suppose the frequency of the emitted thermal radiation is $\omega_0$. Then
\begin{equation}
    \label{H5}
    \omega_0=M_{n+1}-M_n=\frac{M_\text{P}}{\sqrt{2n}}=\frac{M^2_\text{P}}{M}.
\end{equation}
The BH entropy can be expressed in terms of the following adiabatic invariant
\begin{equation}
    \label{H6}
    S_\text{BH}=8\pi\int_{M_\text{P}}^M\frac{dM}{\omega_0}=\frac{A}{4G}=4\pi M^2G,
\end{equation}
where $A=4\pi R_\text{S}^2$ is the BH horizon area, and $R_\text{S}=2MG$ is the Schwarzschild radius. In quantum mechanics, the adiabatic invariance principle related to the Hamiltonian with a discrete spectrum can be explained with the following reasoning. During an adiabatic transformation, any external perturbations applied to the system should have much lower frequencies than the characteristic frequency $\omega_0$. To understand how the adiabatic invariance concept is applied in BH spectroscopy, kindly refer to the literature sources mentioned as Refs. \cite{Kunstatter:2002pj,Liu:2012zzl,Jiang:2012dm}.
  This results in a significant decrease in transitions between the states with successive quantum numbers.
Because this spectrum is equally spaced, the feasible values for the area of a massive BH are equally spaced as well.

One can obtain the fractional extension of the Wheeler--DeWitt equation (\ref{H3}) by replacing the ordinary derivative by Riesz fractional derivative (\ref{Riesz-d'Alembertian}) \cite{Jalalzadeh:2021gtq}
\begin{equation}
    \label{H7}
    -\frac{d^2}{dz^2}\rightarrow \frac{1}{M_\text{P}^{\alpha-2}}\left(-\frac{d^2}{dz^2}\right)^\frac{\alpha}{2},
\end{equation}
where a new minisuperspace coordinate, $z$, defined by $z=x-M/M_\text{P}^2$. 
To obtain the semi-classical solution of this fractional Wheeler--DeWitt equation, we use the alternative equivalent representation of the Riesz fractional derivative \ref{Riesz-d'Alembertian} (or fractional Laplacian), given by 
\begin{equation}
    \label{sh5}
    -\left(-\frac{{d}^2}{{d}z^2}\right)^\frac{\alpha}{2}\psi(z)=c_{1,\alpha}\displaystyle\int_0^\infty\frac{\psi(z-v)-2\psi(z)+\psi(z+v)}{v^{\alpha+1}}{ d}v,
\end{equation}
where $c_{1,\alpha}=\frac{\alpha 2^{\alpha-1}}{\sqrt{\pi}}\frac{\Gamma((1+\alpha)/2)}{\Gamma((2-\alpha)/2)}$ \cite{pozrikidis2018fractional}.

It is a relatively uncomplicated procedure to showcase that the aforementioned fractional operator is indeed non-local in nature. In order to achieve this objective, we shall initiate by scrutinizing the definition of a particle that is localized in Newtonian mechanics. As per the foundational principles of Newtonian mechanics, the magnitude of the alteration in the position of a singular point mass, $x(t)$, with respect to the variable of time, $t$, is commensurate with its velocity, which is epitomized by the primary derivative. Following the works of Weierstraß, the initial c can be defined as 
\begin{equation}\label{derivative}
 v(t)=\frac{dx(t)}{dt}=   \begin{cases}
        \displaystyle\lim_{h\rightarrow0}\frac{x(t)-x(t-h)}{h},&\text{left, causal},\\
\displaystyle\lim_{h\rightarrow0}\frac{x(t+h)-x(t-h)}{2h},&\text{symmetric},\\
\displaystyle\lim_{h\rightarrow0}\frac{x(t+h)-x(t)}{h},&\text{right, anti-causal},\\
    \end{cases}
\end{equation}
where $h\geq0$ and $h<\epsilon$ for an arbitrary small positive real-valued $\epsilon$. As long as the function $x(t)$ is analytic and smooth with respect to $t$, the three definitions provided appear interchangeable. If $t$ is considered a space-like coordinate, the definitions suggest that positional data is collected only from the left or right or concurrently from both directions. If $t$ is regarded as a time-like coordinate, the initial definition in (\ref{derivative}) adheres to the principle of causality, which broadly posits that the current state of a physical object can only be affected by past events. The final definition (\ref{derivative}) conspicuously contravenes the principle of causality in the context of classical mechanics. Notwithstanding, one could embrace Stuckelberg's and Feynman's perspectives in quantum mechanics and recollect their notion that anti-particles travel in reverse time. Consequently, this definition proves fitting to delineate the velocity of an anti-particle. The second definition of a derivative in Eq. (\ref{derivative}), which combines causal and anti-causal propagation, ultimately suggests that relying solely on individual particle interpretation may pose challenges. Hence, examining whether this definition could be used to depict a particle-antiparticle pair's velocity is crucial. Although this discussion may sound quite sophisticated and artificial when it comes to ordinary derivatives, it is important to note that regardless of the definition used, a violation of the causality principle is only of order $\epsilon$. However, for non-local quantities, these considerations become crucial.

As a next step, we shall proceed to redefine the conventional derivative in the context of an integral formulation:
\begin{equation}\label{derivative2}
 v(t)=\frac{d}{dt}\Big|_\text{non local}x(t)=  \begin{cases}
\displaystyle\frac{1}{c} \int_0^\infty W(\sigma,y)v(t-y)dy,\hspace{1.6cm}\text{left, causal},\\
\displaystyle\frac{1}{c}\int_0^\infty W(\sigma,y)\frac{v(t+y)+v(t-y)}{2}dy,\,\,\,\text{symmetric},\\
\displaystyle\frac{1}{c} \int_0^\infty W(\sigma,y)v(t+y)dy,\hspace{.8cm}\text{right, anti-causal},
    \end{cases}
\end{equation}
where $\sigma>0$ is a parameter and
\begin{equation}
    \lim_{\sigma\rightarrow0}W(\sigma,y)=\delta(y).
\end{equation}
One can consider the relations as extensions from the local operator $d/dt$ to a non-local version of the same operator. A potential extension of the above method for any local operator $O_\text{local}(x)$ to its corresponding non-local representation $O_\text{nonlocal}(x)$ is given by \cite{herrmann2011fractional}
\begin{equation}
    O_\text{nonlocal}(x)f(x)=\frac{1}{C} \int_0^\infty \Omega(\sigma,y)U(y)O_\text{local}(x)f(x)dy,
\end{equation}
where $U(y)$ is a shift operator. Choosing $W(\sigma,y)=y^{\sigma-1}$ (allowing for a weakly
singular weight function), $C=\Gamma(\sigma)$, and the shift operator by $U(y)f(x)=f(x+y)-f(x-y)$ we obtain the Riesz fractional derivative (\ref{sh5}) \cite{herrmann2011fractional}. Therefore, the Riesz fractional derivative is a non-local operator as a result
of antisymmetric shift operation. Non-locality is a phenomenon that can be observed in all fractional derivatives and integrals. Temporal processes with time-fractional derivatives are usually called memory. In contrast, spatial-fractional derivatives are associated with large quantum jumps \cite{herrmann2011fractional}. Note that the external time coordinate disappears in the Wheeler--DeWitt equation, and the space-fractional derivative may play a crucial role.

It is worth noting that the presence of Planck mass in the preceding formulation of the Riesz fractional derivative indicates the reality of its quantum gravity roots. This gives us the following fractional Wheeler--DeWitt equation 
\begin{equation}
    \label{H8}
  \frac{1}{2M_\text{P}^{\alpha-1}}\left(-\frac{d^2}{dx^2}\right)^\frac{\alpha}{2}\psi(z)+\frac{M_\text{P}^3}{2}z^2\psi(z)=\frac{M^2}{2M_\text{P}}\psi(z).  
\end{equation}
Note that in a particular case where $\alpha=2$, equation \eqref{H8} reduces to its standard counterpart \eqref{H3}.
The minisuperspace of the aforementioned model is unidimensional, implying that there is only one degree of freedom. The fractional extension of the corresponding model can be attained by substituting the ordinary derivative with its fractional extension. This concept is explained thoroughly in Ref. \cite{Jalalzadeh:2021gtq}. For a thorough understanding of the general formalism of fractional classical and quantum gravity, the interested reader may find more information in Refs. \cite{Calcagni:2013yqa,Calcagni:2021aap,Calcagni:2016azd,Calcagni:2021ipd,Jalalzadeh:2020bqu}.

For the aforementioned fractional Wheeler--DeWitt equation, there is currently no general solution that explicitly bears a dependence on $\alpha$. As a result, the Bohr--Sommerfeld quantization rule may be used. By incorporating the exponential form of the wavefunction, $\psi(z)=\exp(-iS)$ into the fractional Laplacian (\ref{sh5}) and expanding $\psi(z\pm\nu)$ through a Taylor series centered on a, we can leverage the semi-classical approximation to arrive at our final result
\begin{equation}
    \label{H8b}
  -\left(-\frac{d^2}{dx^2}\right)^\frac{\alpha}{2}\psi(z)=c_{1,\alpha}e^{-iS}\int_0^\infty\frac{\sin^2(\frac{\nu S'}{2})}{v^{1+\alpha}}d\nu=|S'|^\alpha\psi(z),
\end{equation}
where $S'=dS/dz$.
Using this relation, and $S'=p$ in Eq. (\ref{H8}) gives us
\begin{equation}
    \label{H9}
   |p|^\alpha+M_\text{P}^{\alpha+2}z^2=M^2M_\text{P}^{\alpha-2},
\end{equation}
where $p$ is the canonical momentum corresponding to $z$.
This equation is the fractional extension of the BH Hamiltonian (\ref{H2}). The classical turning points, i.e., $|p|=0$, are $z=\pm M/M_\text{P}$. Therefore, the Bohr--Sommerfeld quantization rule reads
\begin{multline}
    \label{H10}
    2\pi n=\oint pdz=4\left(\frac{M}{M_\text{P}} \right)^{\frac{2}{\alpha}+1}\int_0^1(1-y^2)^\frac{1}{\alpha}dy=\frac{2\sqrt{\pi}\Gamma(\frac{d+1}{2})}{\Gamma(\frac{d+2}{2})}\left(\frac{M}{M_\text{P}}\right)^d=2\frac{\Omega_d}{\Omega_{d-1}} \left(\frac{M}{M_\text{P}}\right)^d,
\end{multline}
where $\Omega_d$ is the volume of a $d$-dimensional unit sphere, and we defined
\begin{equation}
    \label{H11}
    d=\frac{2}{\alpha}+1.
\end{equation}
The semi-classical mass spectrum below is the result of using the conventional Bohr--Sommerfeld quantization procedure in this situation
\begin{equation}
    \label{H12}
    M=\left(\frac{\Omega_{d-1}}{\Omega_d}n\pi\right)^\frac{1}{d}M_\text{P},~~~~n=\{\text{large positive integer}\}.
\end{equation}
The frequency of the emitted
thermal radiation, the fractional extension of (\ref{H5}), will be
\begin{equation}\label{H12b}
  \omega_0(d)=M_{n+1}-M_n=\frac{\pi\Omega_{d-1}}{d\Omega_d}\left(\frac{M}{M_\text{P}}\right)^{1-d}M_\text{P}.  
\end{equation}
As it is shown in Ref. \cite{Jalalzadeh:2021gtq}, the above fractional spectrum of the BH leads us to the following modified entropy 
\begin{equation}
    \label{H13}
    S_\text{fractal-BH}=S_\text{BH}^\frac{d}{2},
\end{equation}
where $S_\text{BH}=4\pi GM^2$ is the entropy of ordinary BH obtained in Eq. (\ref{H6}). One may derive the aforementioned entropy by employing the adiabatic invariant (\ref{H6}). In such an instance, the emitted thermal radiation's frequency is determined by (\ref{H12b}), thus affording us
\begin{equation}
    \int_{M_\text{P}}^M\frac{dM}{\omega_0(d)}=\frac{\Omega_d}{\pi\Omega_{d-1}(4\pi)^\frac{d}{2}}S_\text{fractal-BH}.
\end{equation}

In addition, the Hawking temperature, denoted as $T_\text{H}$, of the black hole can be obtained from the differential form of the aforementioned adiabatic invariant. To derive $T_\text{H}$, we simply rearrange the equation presented above as
\begin{equation}
    dM=\frac{\Omega_d\omega_0(d)}{\pi\Omega_{d-1}(4\pi)^\frac{d}{2}}dS_\text{fractal-BH}.
\end{equation}
Comparing this relation with the first law of thermodynamics of BHs, $dM=T_\text{H}dS$, gives us
\begin{equation}
    T_\text{H}=\frac{1}{d(4\pi)^\frac{d}{2}}\left(\frac{M}{M_\text{P}}\right)^{1-d}M_\text{P},
\end{equation}
which is the same as the temperature obtained in Ref. \cite{Jalalzadeh:2021gtq}.

It is widely acknowledged that fractional calculus can modify the fractal dimension \cite{tatom1995}. In references \cite{Jalalzadeh:2022uhl,Jalalzadeh:2021gtq}, the authors demonstrate that Lévy's fractional parameter $\alpha$ signifies the fractal dimension of the BH horizon and the cosmological horizon. Additionally, the authors of reference \cite{Rasouli:2022bug} indicate that due to big jumps in minisuperspace, the initial value of the emerged classical scale factor is reliant on the universe's global geometry (or topology). Moreover, they show that the acceleration of the universe and the e-folding of the inflation epoch are direct outcomes of its topology. This implies that the early inflationary epoch is a direct consequence of fractional quantum cosmology. 
In the following section, we will provide a brief overview of the fundamental concepts and definitions of fractal geometry to showcase the impact of fractional Wheeler--DeWitt on the entropy of the BH and elucidate the significance of Eqs. (\ref{H11}) and (\ref{H13}). 
In section \ref{FSS}, we shall employ the notions discussed in section \ref{FF} to exemplify that the fractal dimension of the fractal-fractional BH horizon surface is equal to $d=2/\alpha+1$.

\section{Fractals and fractal dimension}\label{FF}

Before we get into how fractional quantum gravity transforms the smooth structure of a Schwarzschild BH's horizon into a fractal structure, let us review some fractal basics.

Since the surface-to-volume ratio is inversely proportional to the linear size of the system, which is defined by a large number of relevant units, the surface-to-volume ratio for typical macroscopic entities (sphere, cube, etc.) is small. Objects which have had a high surface-to-volume ratio for quite a while are, therefore, porous or hairy. For instance, the lung's high surface-to-volume ratio can be explained by the necessity for fast gas exchange. The respiratory surface of the human lungs (measured with a resolution of 100 $\mu$m) is the size of a tennis court, yet the volume it encloses is just a few liters. B. Mandelbrot \cite{mandelbrot1983} realized the general significance of such systems. He also created the term ``fractal'' and devised a new type of geometry to describe them mathematically.

\begin{figure}
    \centering
    \includegraphics[width=11cm]{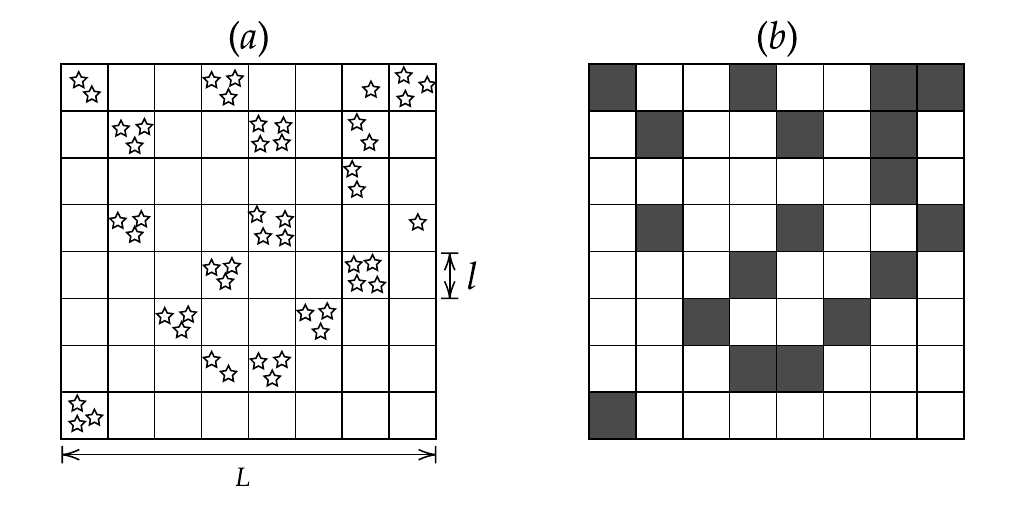}
    \caption{(a) a set (of stars) and a grid of size $l$. $L$ stands for the set's diameter. (b) The black boxes are required to enclose the set.}
    \label{fig1}
\end{figure}

Suppose $D$ is the geometric entity's Euclidean dimension, containing the interests set. The total volume of the boxes necessary to cover the item, please see Fig. \ref{fig1}, or of boxes carrying a portion of the set, is the observed volume $\Omega(l)$ for a given grid of $D$-dimensional cubes of size $l$. A fractal object is one whose perceived volume displays power law behavior with a nontrivial exponent, and resolution (grid size) changes over several orders of magnitude.

It makes natural sense to consider a hairy surface to be an object with dimensions greater than 2. Larger sizes should be used for more ramified surfaces. The idea of the fractal dimension provides a quantitative formulation for this idea. For the sake of this inquiry, assume $L$ to be the set's characteristic linear size. The number of boxes necessary to completely cover the set is denoted by $N(l,L)$ using the previously mentioned box-size $l$ grid (please see Fig. \ref{fig1}). This number can only be determined by a dimensionless quantity, and it must be
\begin{equation}
    \label{F1}
    \varepsilon=\frac{l}{L}.
\end{equation}
This means that the box size is represented in $L$ units. As a result, $N(l, L) = N(\varepsilon)$. With decreasing box size, the number of nonempty cubes grows. Then, the following relation
\begin{equation}
    \label{F2}
 N(\varepsilon)=k\varepsilon^{-d},~~~~\varepsilon\ll 1,   
\end{equation}
specifies the so-called fractal dimension, $d$, as a positive number. Note that the proportionality constant, $k$, is
independent of the resolution. Moreover, $d\leq D$ is attained since $N(\varepsilon)$ cannot be greater than the number of cubes required to fill the space completely.

Fractals are objects that are self-similar or have a common appearance over a broad range of scales in the region defined by Eq. (\ref{F2}). This corresponds to the following scale form description
\begin{equation}
    \label{F3}
    N(\lambda\varepsilon)=k\lambda^{-d}\varepsilon^{-d}=\lambda^{-d}N(\varepsilon),~~~~\lambda>0.
\end{equation}

Fractals are classified into two types: ``deterministic fractals'' and ``random fractals''. In the first class (for example, consider Koch's Curve or Snowflake Fractal), we utilize deterministic rules to construct them. Because the second class is formed by nondeterministic rules, we refer to them as random fractals. The horizon of a BH cannot have a deterministic fractal structure due to the quantum randomness of quantum space-time \cite{Kroger:2000wa}. Several authors \cite{Barrow:2020tzx,Dabrowski:2020atl}, influenced by the geometrical structure of the COVID-19 virus, suggested that the horizon of a BH should have a Koch snowflake fractal shape. This notion clearly contradicts quantum randomness in quantum gravity.

Thus, we are interested in random fractals in this study.
We assume that the fractal geometry of medium $\mathcal B$ under investigation is random. In other words, $\mathcal B$ is considered to be a collection of all realizations $B(\sigma)$ parametrized by elementary events, $\omega$, of the space $\mathcal V$
\begin{equation}
    \mathcal B=\{B(\sigma)|~\sigma\in\mathcal V\}.
\end{equation}
Any realization of form $B(\sigma)$ complies with the laws of quantum physics.

\section{Fractal structure of fractional Schwarzschild BH}\label{FSS}

Let us examine how the preceding section may be applied to the fractional entropy of a BH obtained in Eq. (\ref{H13}), in section \ref{Horizon}. According to Bekenstein and Hawking, the entropy of a BH is equal to the surface area of the BH's horizon in Planck units, $A_\text{P}=4L_\text{P}^2$. Thus, one can rewrite the Eq. (\ref{H13}) as
\begin{equation}
    \label{FB1}
    S_\text{fractal-BH}=\frac{A_\text{fractal}}{A_\text{P}}.
\end{equation}
In this equation, we defined
\begin{equation}
    \label{FB2}
    A_\text{fractal}=A_\text{P}\left(\frac{A_\text{S}}{A_\text{P}}\right)^\frac{d}{2}=4\pi^\frac{d}{2}L_\text{P}^{2-d}R_\text{S}^d,
\end{equation}
where $A_\text{S}=16\pi G^2M^2$ is the Schwarzschild area of the BH, with corresponding radius $R_\text{S}=2MG$. According to definition (\ref{F2}), the number of boxes (squares) necessary to cover the surface area of the fractional BH
\begin{equation}
    \label{FB3}
    N=\frac{A_\text{fractal}}{A_\text{P}}=(4\pi)^\frac{d}{2}\varepsilon^{-d},
\end{equation}
where $\varepsilon=M_\text{P}/M\ll1$. This shows that the entropy is equal to the number of squares to cover the surface area of the horizon, $S_\text{fractal-BH}=N$ completely. In addition, regarding the definition (\ref{F2}), the horizon of fractional BH is a fractal surface with a dimension equal to $d=2/\alpha+1$, as defined in Eq.(\ref{H11}).
Regarding Lévy’s fractional parameter, $\alpha$  is restricted to the interval $1 <\alpha \leq 2$ (please see, for example, \cite{Laskin:2002zz}), the fractal dimension of fractal-fractional BH is given by the following interval
\begin{equation}
    \label{FB4}
    2\leq d<3.
\end{equation}

Employing the fractal BH entropy (\ref{FB1}), one can define effective Schwarzschild radius, $R_\text{eff}$ and mass, $M_\text{eff}$ as
\begin{equation}\label{FB5}
    \begin{split}
   R_\text{eff}&=(4\pi)^\frac{d-2}{4}\left(\frac{R_\text{S}}{L_\text{P}}\right)^\frac{d}{2}L_\text{P}.\\  
   M_\text{eff}&=\frac{R_\text{eff}}{2G}=(4\pi)^\frac{d-2}{4}\left(\frac{M}{M_\text{P}}\right)^\frac{d}{2}M_\text{P}.
    \end{split}
\end{equation}
As a result, one can rewrite the fractal surface area and fractal entropy of the BH as
\begin{equation}
    \label{FB6}
    \begin{split}
        A_\text{fractal}&=4\pi R^2_\text{eff},\\
        S_\text{fractal-BH}&=\frac{A_\text{fractal}}{4G}=\frac{\pi}{G}R_\text{eff}^2.
    \end{split}
\end{equation}
Now, the generic expression for the fractional-fractal Bekenstein--Hawking temperature may be deduced using the first law of thermodynamics
\begin{equation}
    \label{FB7}
    \frac{1}{T_\text{eff}}=\frac{dS_\text{fractal-BH}}{dM_\text{eff}},~~~~~T_\text{eff}=\frac{1}{2\pi R_\text{eff}}.
\end{equation}
Note that we use units in which $\hbar=c=k_B=1$.

It is worth noting that when $d=2$, or equivalently $\alpha=2$, all fractional-fractal quantities acquired in the preceding will be reduced to their original values.

\section{Emergent Cosmology}

According to Padmanabhan's proposal \cite{Padmanabhan:2012ik}, classical gravity is an emerging phenomenon, and cosmic space emerged as cosmic time advanced. He claimed that the difference in the number of degrees of freedom between the holographic surface and the emerging bulk is proportional to the change in the cosmic volume. In this sense, he could effectively extrapolate the Friedmann equation describing the Universe's evolution with zero spatial curvature.
Furthermore, claiming that space emerges around finite gravitational systems, such as the Sun-Earth system, is not the wisest course of action. Nevertheless, Padmanabhan intriguingly demonstrated that this issue is moot, at least in the context of cosmology, by picking any chosen time interval, say cosmic time. This is how Padmanabhan came to make the claim that the Universe will keep expanding until holographic equipartition takes place. He postulated that the variation of the cosmic volume, $dV$, in an infinitesimal interval of cosmic time, $dt$, is given by
\begin{equation}
    \label{P1}
    \frac{dV}{dt}=L_\text{P}^2(N_\text{sur}-N_\text{bulk}),
\end{equation}
where $N_\text{sur}$ and $N_\text{bulk}$ are the surface and bulk degrees of freedoms, respectively. For a flat ($k=0$) FLRW Universe, as Padmanabhan assumed \cite{Padmanabhan:2012ik}, the cosmic volume, the surface degree of freedom, and the bulk degree of freedom are 
\begin{equation}
    \label{P2}
    V=\frac{4\pi}{3}R_{H}^3,~~~~N_\text{sur}=\frac{4\pi R^2_\text{H}}{L_\text{P}^2},~~~~N_\text{bulk}=\frac{2|E_\text{Komar}|}{T},
\end{equation}
where $R_\text{H}$ defined by
\begin{equation}
    \label{P3}
    R_\text{H}=\frac{1}{H},\hspace{.5cm}(H=\frac{\dot a}{a}),
\end{equation}
is the radius of the Hubble horizon, $V$ is its volume, 
\begin{equation}
    \label{P4}
    T=\frac{1}{2\pi R_\text{H}}=\frac{H}{2\pi},
\end{equation}
is the temperature of the horizon, and
\begin{equation}
    \label{P5}
    |E_\text{Komar}|=\epsilon(\rho+3p)V,~~~\epsilon=\begin{cases}
    +1,~~~\text{if}~~\frac{p}{\rho}<-\frac{1}{3},\\
     -1,~~~\text{if}~~\frac{p}{\rho}>-\frac{1}{3},
\end{cases}
\end{equation}
is the proper Komar energy of a perfect fluid ($\rho$ is the energy density, and $p$ is the pressure of the fluid) contained inside the Hubble volume. Substituting Eqs. (\ref{P2}) into (\ref{P1}) gives the  Raychaudhuri equation
\begin{equation}
    \frac{\ddot a}{a}=-\frac{4\pi G}{3}(\rho+3p).
\end{equation}
Note that to obtain the Friedmann equation, one also needs the continuity equation $\dot \rho=-3H(\rho+p)$. It is well known the continuity equation is equivalent to
\begin{equation}
    \label{P5a}
    \frac{dM_\text{MSH}}{dt}=-p\frac{dV_\text{PV}}{dt}, 
\end{equation}
where $M_\text{MSH}$ is the Misner--Sharp--Hernandez mass,
\begin{equation}
    \label{P5b}
    M_\text{MSH}=\rho V_\text{PV}.
\end{equation}
In these equations, $V_\text{PV}$ is the areal (proper) volume  
\begin{equation}
    \label{P5c}
    V_\text{PV}=\frac{4\pi}{3} R_\text{PR}^3
\end{equation}
and $R_\text{PR}=ra(t)$ is the proper, or areal, radius.

\section{Modified emergent Cosmology from fractional-fractal entropy}

This paper aims to derive the modified Friedmann and Raychaudhuri equations from the BH's fractional-fractal entropy of cosmic space. Inspired by the expression of fractal entropy (\ref{FB6}), assume that the effective area of the apparent horizon, which serves as our holographic screen, is a random fractal surface similar to the fractal BH obtained in the preceding section. The conclusions found in Reference \cite{Jalalzadeh:2022uhl} for a fractional quantum gravity of de Sitter space provide validity for this assumption. Their findings reveal that the effective area of the de Sitter space's apparent classical horizon has the same random fractal structure as the equation (\ref{FB6}). As a result, extending the fractional-fractal surface of the apparent horizon is natural to a general homogeneous and isotropic Universe. Thus, let us define the effective random fractal radius, $R_\text{eff}$, the surface area $A_\text{fractal-H}$, and the corresponding volume, $V_\text{fractal-H}$ of the Hubble horizon by
\begin{equation}
    \label{P6}
    \begin{split}
    & R_\text{eff}=(4\pi)^\frac{d-2}{4}\left(\frac{R_\text{H}}{L_\text{P}}\right)^\frac{d}{2}L_\text{P},\\
  & A_\text{fractal-H}=4\pi R_\text{eff}^2,\\
   &V_\text{fractal-H}=\frac{4\pi}{3}R_\text{eff}^3.
\end{split}
\end{equation}

\noindent
In addition, we use the modified version of (\ref{P1}) given by
\begin{equation}\label{P7}
     \frac{dV_\text{fractal-H}}{dt}=\frac{d}{2}L_\text{P}^2R_\text{eff}H(N_\text{sur}-N_\text{bulk}),
\end{equation}
proposed in Ref. \cite{Hashemi:2013oia}. This extension can produce the Friedmann and Raychaudhuri equations in higher-order gravity theories, such as Gauss--Bonnet and Lovelock gravities, with any spatial curvature. Using relations (\ref{P6}) in Eqs. (\ref{P5}) and (\ref{P2}) one can obtain the $N_\text{sur}$ and $N_\text{bulk}$
for their fractal extensions
\begin{equation}\label{P2a}
   \begin{split}
  &N_\text{sur}=\frac{4\pi R^2_\text{eff}}{L_\text{P}^2},~~~~N_\text{bulk}=\frac{2|E_\text{Komar}|}{T},\\
&  T=\frac{1}{2\pi R_\text{eff}},~~~|E_\text{Komar}|=\epsilon\sum_i(\rho_i+3p_i)V_\text{fractal-H},      
   \end{split}
\end{equation}
where in the Komar energy, we consider the mixture of perfect fluids with $\rho_i=\omega_ip_i$.
Substituting resulting relations into (\ref{P7}) gives us
\begin{equation}
    \label{P8}
3\frac{\dot R_\text{eff}}{ R_\text{eff}}=\frac{3dH}{2}-2\pi dL_\text{P}^2\epsilon\sum_i(\rho_i+3p_i)H R_\text{eff}^2.  
\end{equation}
where an overdot means a time derivative. 
Using the definition of $ R_\text{eff}$ in Eq. (\ref{P6}) one can simplify the above equation into
\begin{equation}
    \label{P8a}
    \dot H+H^2=-\frac{4\pi L_\text{P}^2}{3}\sum_i(\rho_i+3p_i)H^2 R_\text{eff}^2,
\end{equation}
which is the fractional-fractal extension of the Raychaudhuri equation. To obtain the corresponding Friedmann equation, first, we have to obtain the fractional-fractal extension of the continuity equation (\ref{P5a}). To this end, we define the fractional-fractal extension of the proper distance and proper volumes (\ref{P5b}) and (\ref{P5c})
\begin{equation}
\begin{split}
     \label{P8b}
 &  R_\text{eff-PD}=(4\pi)^\frac{d-2}{4}\left(\frac{R_\text{PD}}{L_\text{P}}\right)^\frac{d}{2}L_\text{P},\\ &V_\text{fractal-PV}=\frac{4\pi}{3}R_\text{eff-PD}^3.
\end{split}
   \end{equation}
As a result, the effective Misner--Sharp--Hernandez mass will be 
\begin{equation}\label{P8c}
    M^\text{(eff)}_{MSH}=\sum_i\rho_i V_\text{fractal-PV}.
\end{equation}
The continuity equation (\ref{P5a}) modify to
\begin{equation}
    \label{P8d}
     dM_\text{MSH}^\text{(eff)}=-\sum_ip_idV_\text{fractal-PD}.
\end{equation}
This equation is equivalent to 
\begin{equation}
    \label{P8e}
    \dot\rho_i=-\frac{3d}{2}(\rho_i+p_i)H.
\end{equation}
In addition, note that the above extension of the proper distance gives us the fractional extension of the redshift-scale factor relation. The scale factor is related to the observed redshift $z$ of the light emitted at the time 
$t_\text{em}$ by
\begin{equation}
    \label{redshift}
    \left(\frac{a_0}{a(t_\text{em})}\right)^\frac{d}{2}=1+z.
\end{equation}

Using the modified continuity equation (\ref{P8e}), one can show that
\begin{equation}
    \label{P8f}
    (\rho_i+3p_i)H=-\frac{2}{da^d}\frac{d}{dt}(a^d\rho_i).
\end{equation}
Now, using the definition of effective radius, $R_\text{eff}$, in (\ref{P6}), and  the above equation, one can simplify Eq. (\ref{P8}) into
\begin{equation}
    \label{P9}
    \frac{d}{dt}\left(a^dH^d\right)=\frac{2}{3}(4\pi)^\frac{d}{2}L_\text{P}^{4-d}\frac{d}{dt}\sum_i(a^d\rho_i).
\end{equation}
This gives us the fractional-fractal extension of the Friedmann equation for the flat space
\begin{equation}
    \label{P10}
    H^d=\frac{2}{3}(4\pi)^\frac{d}{2}L_\text{P}^{4-d}\sum_i\rho_i.
\end{equation}
One can rewrite the above equation as
\begin{equation}
    \label{P10a}
   H^2=\frac{8\pi G}{3}\left(\frac{2}{3\rho_\text{P}}\sum_j\rho_j\right)^{\frac{2}{d}-1}\sum_i\rho_i,
\end{equation}
where $\rho_\text{P}=1/L_\text{P}^4=5.1550\times10^{96}$ Kg/m$^3$ is the Planck energy density.
Again, regarding the definition of $R_\text{eff}$, using the relations $\dot H=\ddot a/a-H^2$ and (\ref{P10}) in (\ref{P8a}) we obtain the fractional-fractal Raychaudhuri equation
\begin{equation}
    \label{P11}
 \frac{\ddot a}{a}=-\frac{4\pi G}{3}\left(\frac{2}{3\rho_\text{P}}\sum_j\rho_j\right)^{\frac{2}{d}-1}\sum_i(\rho_i+3p_i).
\end{equation}
It is worth noting that when $d=2$ (or $\alpha=2$), Eqs. (\ref{P10a}) and (\ref{P11}) will be reduced to the original Friedmann and Raychaudhuri equations.

\section{Lambda-CDM cosmological model}

In this part, we briefly review some basic aspects of the standard model of cosmology in preparation for the modification and extension of the $\Lambda$CDM model in the following section.

The $\Lambda$CDM model is a formulation of the relativistic cosmology, according to which the Universe is made up of three main components: ordinary matter, the postulated cold dark matter (CDM), and a cosmological constant, $\Lambda$. Due to its simplicity and ability to fairly explain the cosmic microwave background's existence and structure the large-scale structure of the Universe, the observed abundances of lithium, helium, and hydrogen (including deuterium), and the late time acceleration of the Universe, it is commonly referred to as the standard model of cosmology. In this model, which has been used in observational cosmology since the 1960s, the Friedmann equation gives the time evolution of the scale factor as a function of the Hubble parameter at the present time, $H_0$, and three independent density parameters. The Friedmann equation for a spatially flat Universe of the standard model of cosmology is
\begin{equation}
    \label{S00}
    H^2=H_0^2\left\{\Omega_0^\text{(rad)}\left(\frac{a_0}{a}\right)^{4}+ \Omega_0^\text{(cm)}\left(\frac{a_0}{a}\right)^{3}+\Omega^{(\Lambda)}\right\},
\end{equation}
 where $\Omega_0^\text{(i)}=\frac{8\pi G\rho_i(t_0)}{3H_0^2}$ are density parameters of radiation, $i=$``rad'', cold matter, $i=$``cm'' and the cosmological constant, $i=\Lambda$, at the present time, $t=t_0$, respectively. 

The Hubble parameter $H_0$ is usually written as
\begin{equation}
    \label{Hubble}
H_0=100h~\text{Km}~\text{sec}^{-1}~\text{Mpc}^{-1}=2.1332h \times 10^{-42}~ \text{GeV},
\end{equation}
where $h$ represents the uncertainty on the value $H_0$. The observations of the Planck 2018 collaboration \cite{Planck:2018vyg} constrain this value to be $h=0.674\pm0.005$ based on \textit{Planck} TT, TE, EE+lowE+lensing CMB data. By Planck's 2018 collaboration, the current dark energy density parameter is
\begin{equation}
    \label{Dark}
 \Omega_0^{(\Lambda)}=0.685 \pm 0.007.  
\end{equation}
 The density parameter $\Omega^\text{(cm)}_0$ is equal to the total of the baryon $\Omega^\text{(b)}_0$ and CDM $\Omega^\text{(CDM)}_0$ contributions, i.e. $\Omega^\text{(cm)}=\Omega^\text{(CDM)}_0+\Omega^\text{(b)}_0$. Following Planck 2018 collaboration \cite{Planck:2018vyg}, for $\Omega^\text{(b)}_0$ and $\Omega^\text{(CDM)}_0$, respectively we have
 \begin{equation}
     \label{5-6}
     \begin{split}
\Omega^\text{(b)}_0h^2&=0.02237 \pm 0.00015,\\
\Omega^\text{(CDM)}_0h^2&=0.1200 \pm 0.0012.
\end{split}
\end{equation}
 Assuming $h=0.674$, then the density parameters are $\Omega^\text{(b)}_0=
0.04924319$, and $\Omega^{\text{(CDM)}}_0=0.2641566$, for the central value. 
Thus,
\begin{equation}
    \label{matter}
    \Omega^\text{(cm)}_0=0.315.
\end{equation}
Results from the Planck 2018 collaboration \cite{Planck:2018vyg} for TT, TE, EE+lowE+lensing added with BAO data also suggests that our Universe is accurately spatially flat with a curvature density $\Omega^{(k)}_0 = 0.0007 \pm 0.0019$.

The expression (\ref{S00}) gives us the age of the Universe 
\begin{equation}
    \label{age}
    t_0=\frac{1}{H_0}\int_0^1 \frac{dx}{x\left[\Omega_0^\text{(rad)}x^{-4}+ \Omega_0^\text{(cm)}x^{-3}+\Omega^{(\Lambda)}\right]^\frac{1}{2}},
\end{equation}
where $x=a/a_0$. Since the current energy density parameter of
the radiation is of the order of $10^{-5} - 10^{-4}$, radiation becomes
important only for high redshifts, i.g., $z{\,}{\simeq}{\,}1000$. Hence, it is a reasonable approximation to neglect the contribution from radiation
in the above integral. The cosmic age of the Universe is constrained to be $t_0=13.797\pm 0.023$ Gyr \cite{Planck:2018vyg}. Under this bound, we find that the density parameter of the non-relativistic matter
is constrained to satisfy (\ref{matter}). As a result, CDM is an essential component of the standard model of cosmology in order to explain the age of the Universe.

At the present epoch, the deceleration parameter, $q$, is connected to the combination of $\Omega_0^\text{(cm)}$ and $\Omega^{(\Lambda)}$ by $q=-\Omega^{(\Lambda)}+\Omega_0^\text{(cm)}/2$.

\section{Lambda-Cold Baryonic Matter cosmological model}

In a similar fashion to the standard model of cosmology, we assume the admixture of three species for the matter content of the model Universe: relativistic radiation, with the equation of state parameter $\omega=1/3$, cold matter, with $\omega=0$, and the cosmological constant, $\Lambda$ with $\omega=-1$. Using the continuity equation (\ref{P8e}) gives the energy density of these species as
\begin{equation}
    \label{S1}
    \begin{split}
    \rho_\text{rad}(t)&=\rho_\text{rad}(t_0)\left(\frac{a}{a_0}\right)^{-2d},\\
    \rho_\text{cm}(t)&=\rho_\text{cm}(t_0)\left(\frac{a}{a_0}\right)^{-\frac{3d}{2}},\\
    \rho_\Lambda(t)&=\rho_\Lambda(t_0),
    \end{split}
\end{equation}
where $t_0$ stands for the present epoch, ``rad'' and ``cm'' denote relativistic radiation and cold matter, respectively. Plugging these relations into the Friedmann equation (\ref{P10a}) gives 
\begin{equation}
    \label{S2}
    H^2=H_0^2\Big\{\Omega_0^\text{(rad,fractal)}x^{-2d}+ \Omega_0^\text{(cm,fractal)}x^{-\frac{3d}{2}}+\Omega^{(\Lambda,\text{fractal})}\Big\}^\frac{2}{d},
\end{equation}
where $x=a/a_0$, $H_0$ is the Hubble parameter at the present time, and
\begin{equation}
    \label{S3}
    \Omega_0^\text{(i,fractal)}=\frac{8\pi G\rho_i(t_0)}{3H_0^2}\left(\frac{L_\text{P}H_0}{2\sqrt{\pi}}\right)^{2-d}=\Omega_0^\text{(i)}\left(\frac{L_\text{P}H_0}{2\sqrt{\pi}}\right)^{2-d},
\end{equation}
denote the density parameters of three species and the subscript 
$i$ is one of $i=$rad, cm, $\Lambda$. 
Note that $\Omega_0^\text{(i)}$ denotes the standard density parameters of $\Lambda$CDM cosmology defined in (\ref{S00}).
It follows that similar to the standard cosmology, the present density parameters defined in Eq. (\ref{S3}) obey
\begin{equation}
    \label{S4}
\Omega_0^\text{(rad,fractal)}+\Omega_0^\text{(cm,fractal)}+\Omega_0^{(\Lambda,\text{fractal})}=1.
\end{equation}
Using the fractional-fractal Friedmann equation (\ref{P10a}), we can determine the age of the Universe as
\begin{equation}
    \label{S5}
    t_0=\frac{1}{H_0}\int_0^1 \frac{dx}{x\Big[\Omega_0^\text{(rad,fractal)}x^{-2d}+ \Omega_0^\text{(cm,fractal)}x^{-\frac{3d}{2}}+\Omega^{(\Lambda,\text{fractal})}\Big]^\frac{1}{d}}.
\end{equation}
Table \ref{tabl1} shows the age of the Universe for various values of $d$. Here, we neglect the contribution of radiation, and we assumed $\Omega_0^\text{(cm,fractal)}=0.315$ (similar to the standard $\Lambda$CDM cosmology).
 \begin{table}[h!]
    \centering
    \begin{tabular}{|c|c|c|}
    \hline
      $d$  & $t_0$ (Gyr)& $\Omega_0^\text{(cm)}$ \\
      \hline
      2& 13.797& 0.315\\
      \hline
     2.01314& 13.777& 0.049\\
      \hline
      2.1 & 13.648 & $2.2\times 10^{-7}$\\
      \hline
      2.5 & 13.147 & $5.7\times 10^{-32}$\\
      \hline
      2.7 & 12.941 & $2.9\times 10^{-44}$\\
      \hline
       2.9 & 12.760 & $1.5\times 10^{-56}$\\ [1ex]
           \hline
          \end{tabular}
    \caption{The age of Universe, $t_0$ and the density parameter of cold matter for various values of fractal dimension $d$. Here we consider $h=0.674$ and the Hubble time $t_H=1/H_0=14.508$ Gyr.}
    \label{tabl1}
\end{table}
As can be seen, altering the fractal dimension affects the Universe's age, which ranges from 12.760 Gyr to 13.797 Gyr, and this last value is obtained from $d=2$ in which the standard $\Lambda$CDM cosmology holds. On the other hand, it significantly alters the parameter for the ``actual'' density of the Universe's cold matter composition. According to Eq. (\ref{S3}) the ordinary density parameter is
\begin{equation}
    \label{S3b}
\Omega_0^\text{(cm)}=\frac{8\pi G\rho_m(t_0)}{3H_0^2}=\Omega_0^\text{(cm,fractal)}\left(\frac{L_\text{P}H_0}{2\sqrt{\pi}}\right)^{d-2}.
\end{equation}
Altering $d$ from two to three changes the required cold matter density from 0.315 to $10^{-56}$; please see Table \ref{tabl1}. This means that the fractal dimension could highly amplify the effective contribution of the matter content in the cosmological parameters. A very interesting case is $d=2.01314$. For this value of the fractal dimension, the actual density parameter of the cold matter is $\Omega^\text{(cm)}_0=0.049$, which is equal to the density parameter of baryonic matter $\Omega^\text{(b)}_0$. This means that by assuming the total cold matter content of the Universe that is measured is of fractal origin, the baryonic matter density parameter supersedes the need for a CDM component. Thus, introducing the CDM density parameter as part of the cold matter of the Universe on standard $\Lambda$CDM cosmology is reflected in the current cosmological model as the fractal extension expressed by Eq. (\ref{S3b}) with $d=2.01314$.

In addition, using the fractional-fractal Raychaudhuri equation (\ref{P11}) and the new fractal redshift relation (\ref{redshift}), one obtains the deceleration parameter as 
\begin{equation}\label{S6}
    q(z)=
    \frac{2\Omega_0^\text{(rad,fractal)}(1+z)^4+\Omega_0^\text{(cm,fractal)}(1+z)^3-2\Omega_0^{(\Lambda,\text{fractal})}}{2\Big[\Omega_0^\text{(rad,fractal)}(1+z)^4+\Omega_0^\text{(cm,fractal)}(1+z)^3+\Omega_0^{(\Lambda,\text{fractal})}\Big]}.
\end{equation}
This demonstrates that the deceleration parameter is the same as the conventional one. As a result, the redshift from deceleration to the acceleration phase is identical to the standard model of cosmology.

\section{Conclusions}
According to fractional quantum gravity, a black hole's event horizon is a random fractal surface with a dimension less than three and more than or equal to two. In addition, the event horizon of de Sitter space-time follows the same property. These tips take the concept of the fractal horizon and apply it to the cosmic context. We obtained the Friedmann and Raychaudhuri equations utilizing Padmanabhan's emergent cosmic space paradigm by assuming a fractal horizon surface for an isotropic and homogeneous universe.

Our model provides an alternative perspective, implying that fractional-fractal characteristics may have impacted the measurable features of the Universe's matter content by modifying the density parameter of cold matter. Specifically, altering the fractal dimension could highly amplify the baryonic matter content's effective contribution. Therefore, just as the fractal-fractional alteration prevents the need for dark matter at galactic scales \cite{Giusti:2020rul,Giusti:2020kcv,Varieschi:2020ioh,Varieschi:2020dnd,Varieschi:2020hvp,Calcagni:2021mmj}, it can do the same cosmological scales.

\section*{Acknowledgments}

S.J. acknowledges financial support from the National Council for Scientific and Technological Development--CNPq, Brazil, Grant no. 308131/2022-3.

\bibliographystyle{unsrt}
\bibliography{sn-bibliography}

\end{document}